\documentclass[10pt,prb,aps,superscriptaddress,twocolumn,longbibliography]{revtex4-1}
\usepackage{amssymb,amsmath}
\usepackage{graphicx}
\usepackage{color}
\usepackage{natbib}
\usepackage{hyperref}
\usepackage[utf8]{inputenc}

\begin{document}

\newcommand{\sgn}{\mathop{\mathrm{sgn}}}
\newcommand{\e}{{\rm e}}
\newcommand{\rmi}{{\rm i}}
\renewcommand{\Im}{\mathop\mathrm{Im}\nolimits}
\renewcommand{\Re}{\mathop\mathrm{Re}\nolimits}
\newcommand{\red}[1]{{\color{red}#1}}
\newcommand{\blue}[1]{{\color{blue}#1}}
\newcommand{\yellow}[1]{{\color{yellow}#1}}
\newcommand{\commentMaxim}[1]{{\color{blue}{\it ~Maxim:~}\tt #1}} 

\renewcommand{\cite}[1]{[\onlinecite{#1}]}
\newcommand{\cra}[1]{\hat{a}^{\dag}_{#1}}  
\newcommand{\ana}[1]{\hat{a}_{#1}^{\vphantom{\dag}}}         
\newcommand{\num}[1]{\hat{n}_{#1}}         
\newcommand{\ket}[1]{\left|#1\right>}      
\newcommand{\bra}[1]{\left<#1\right|}
\newcommand{\eps}{\varepsilon}      
\newcommand{\om}{\omega}      
\newcommand{\kap}{\varkappa}      

\newcommand{\spm}[1]{#1^{(\pm)}}
\newcommand{\skvk}[2]{\left<#1\left|\frac{\partial #2}{\partial k}\right.\right>} 
\newcommand{\skvv}[2]{\left<#1\left|#2\right.\right>}
\newcommand{\df}[2]{\frac{\partial #1}{\partial #2}}
\newcommand{\ds}[1]{\partial #1/\partial k}

\title{Interaction-induced topological states of photon pairs}

\author{Andrei A. Stepanenko}
\affiliation{Department of Physics and Engineering, ITMO University, Saint Petersburg 197101, Russia}

\author{Maxim~A.~Gorlach}
\email{m.gorlach@metalab.ifmo.ru}
\affiliation{Department of Physics and Engineering, ITMO University, Saint Petersburg 197101, Russia}

\begin{abstract}
To date, the concept of topological order relies heavily on the properties of single-particle bands. Only recently it has been realized that interactions can have a dramatic impact on topological properties not only modifying the  topology of the bands but also creating a topological order in an otherwise trivial system. Applying an extended version of the Bose-Hubbard model, we investigate a system which, being topologically trivial in the single-particle regime, harbors topologically nontrivial edge and interface states of repulsively bound photon pairs. Whereas binding of the photons in this model is captured by a standard local interaction term, an additional direct two-photon hopping renders the system topologically non-trivial. Besides their interaction-induced origin, predicted two-photon edge states exhibit a range of other unexpected features, including the robustness to collapse of the corresponding bulk band and the ability to coexist with the continuum of two-photon scattering states forming a bound state in the continuum. Performing rigorous calculation of the Zak phase for bound photon pairs, we prove the topological origin of the two-photon edge states.
\end{abstract}

\maketitle

\section{Introduction}\label{sec:Intro}

Topological photonics offers a rich variety of remarkable functionalities including disorder-robust routing of light on a chip~\cite{Lu2014,Lu2016,Khanikaev-review,Ozawa_RMP,Rider2019,Smirnova-toporeview}. While topological states in classical optical systems form an established area of research~\cite{Lu2014,Lu2016,Khanikaev-review,Ozawa_RMP,Rider2019,Smirnova-toporeview}, the emphasis is currently shifting towards topological states of quantum light~\cite{Roushan2014,Barik,Tambasco,Mittal-2018,Blanco-Science,Wang2019} with the potential of applications in topologically protected quantum information transfer, quantum computations and manipulation of entangled photons with quantum metasurfaces~\cite{Sukhorukov-Science}.

Just within one year, first realizations of single-photon topological states~\cite{Barik,Tambasco} and topologically protected sources of non-classical light~\cite{Mittal-2018} have been reported. Moreover, previous theoretical analysis of entangled photons propagation in a topological system~\cite{Rechtsman-Segev,Mittal-Hafezi} has been followed by recent experiments~\cite{Blanco-Science,Wang2019}. In this context, it is especially important to investigate the implications of topological protection for more complex quantum states of light which can potentially uncover further exciting applications of topological photonics.
 
One of such intriguing states of quantum light is represented by doublons, which are bound photon pairs arising in discrete nonlinear arrays due to repulsive Kerr-type nonlinearity~\cite{Mattis1986,Winkler}. Quite counter-intuitive properties of doublon quasi-particles were analyzed in a series of theoretical papers in the context of bulk~\cite{Valiente,Valiente2009,Menotti,Bello-2017,Wang-Liang} and edge doublon states~\cite{Pinto,Flach,Zhang2012,Zhang2013,Longhi,Gorlach-H-2017,DiLiberto-EPJ} including more advanced concepts of doublons in two-dimensional geometries~\cite{Salerno,Salerno2019}, Thouless pumping of doublons~\cite{Angelakis2016,YKe} and dissipatively bound photon pairs~\cite{Lyubarov}.

Driven by the ambitious goal to realize topological doublon edge states, we and several other groups have investigated a well-celebrated Su-Schrieffer-Heeger model (SSH)~\cite{Su} in the two-photon regime with the effective on-site repulsive photon-photon interaction~\cite{DiLiberto,Gorlach-2017,Marques2018}. However, since the analyzed model is topologically nontrivial even in the single-particle case, the emergence of two-photon edge states~\cite{Gorlach-2017}  
it is not so surprising. Therefore, it is much more exciting to demonstrate topological states of doublons induced by interactions in an otherwise topologically trivial system.

Interestingly, such interaction-induced topological states are already known for classical systems characterized by the intensity-dependent coupling constants between some of the sites which give rise to the self-induced topological transitions~\cite{Hadad,Hadad-ACS,Hadad-Nature}. 

\begin{center}
    \begin{figure}[b]
     \includegraphics[width=1\linewidth]{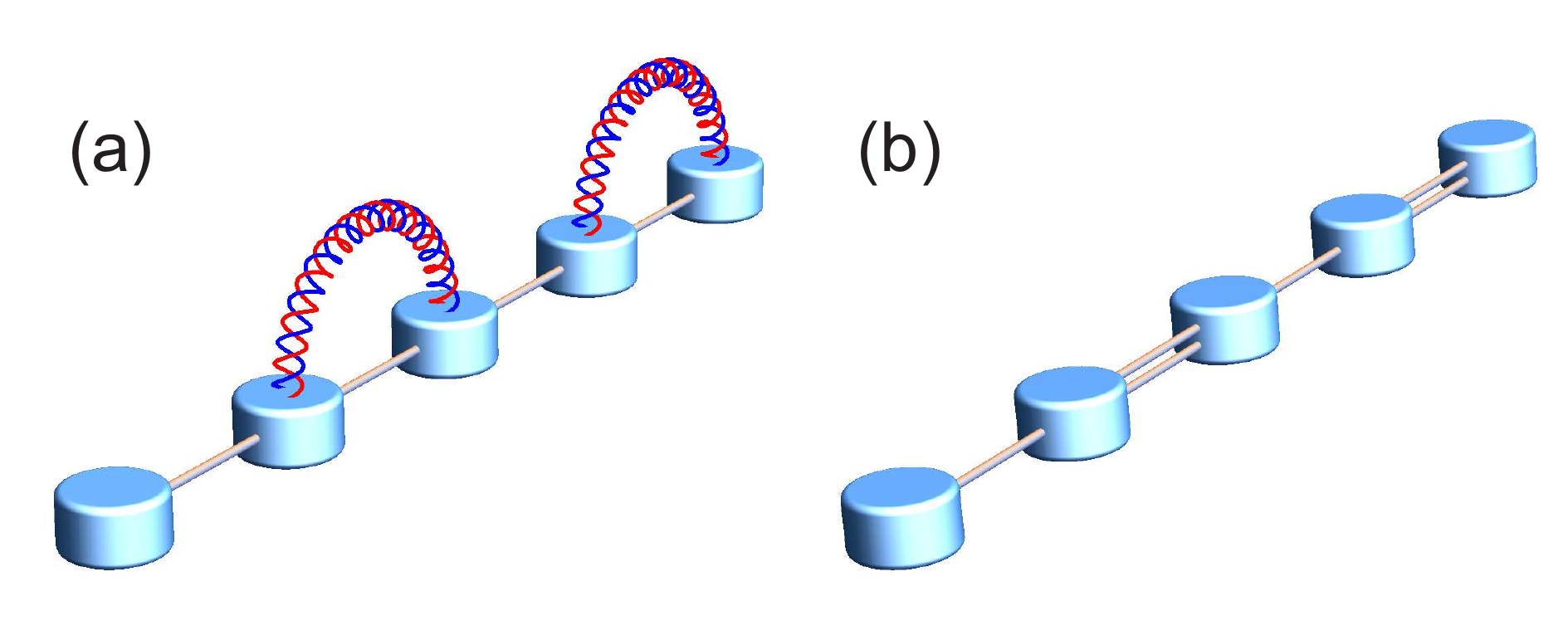}
     \caption{(a) The sketch of the system under study. Straight connecting lines represent single-photon tunneling amplitude $J$, whereas wavy lines illustrate effective two-photon hopping $P$, which enters the extended Bose-Hubbard Hamiltonian Eq.~\eqref{Hamiltonian}. (b) In the limit of strong interactions $U\gg J$ the dynamics of bound photon pair is governed by the Su-Schrieffer-Heeger model.
}
    \label{fig:Sketch}
    \end{figure}
    \end{center}

To demonstrate interaction-induced topological states of photon pairs, we have recently proposed~\cite{Olekhno} a one-dimensional system depicted schematically in Fig.~\ref{fig:Sketch}(a) which, besides local photon-photon interaction also incorporates a direct two-photon hopping, which does not affect single-particle eigenstates and energies but becomes effective in the presence of two photons.

In this Article, we investigate and advance the concept of interaction-induced topological doublon states in the presence of the direct two-photon hopping, deriving the dispersion of bulk doublons and calculating the Zak phase for them. Our results prove the topological origin of the interaction-induced doublon states and provide valuable insights into the problem of topological characterization of few-body states. 

Quite importantly, topological states of doublons studied here should not be mixed with soliton-like nonlinear topological states of classical light arising in waveguide lattices with Kerr-type nonlinearity~\cite{Lumer,Leykam2016} or mean-field solutions of nonlinear Gross-Pitaevskii equation in the form of vortices~\cite{Solnyshkov-NC} because of the few-body nature of topological states in our proposal.

The rest of the paper is organized as follows. In Sec.~\ref{sec:Model} we summarize our model and provide simple arguments to prove the existence of interaction-induced topological states. Section~\ref{sec:Dispersion} contains an  in-depth analysis of the bulk properties of bound photon pairs including an analytical model for their dispersion and diagrams showing the evolution of doublon bands when the parameters of the model are varied. The properties of the edge and interface doublon states are examined in Sec.~\ref{sec:Edge}, whereas our conclusions and outlook for future studies appear in Sec.~\ref{sec:Concl}. Technical details regarding the calculation of bulk doublons dispersion and Zak phase are summarized in Appendices A and B, respectively.

\section{Summary of the model and doublon edge states}\label{sec:Model}

We search for the eigenstates of the system described by the extended version of Bose-Hubbard Hamiltonian:
\begin{eqnarray}
\label{Hamiltonian}
  \hat{H} &=& \omega_0\sum_{m} \hat{n}_m -J \sum_{m}(\cra{m} \ana{m+1}+\cra{m+1} \ana{m})\nonumber\\
    &&+ U \sum_{m} \hat{n}_m(\hat{n}_m-1)\nonumber\\
    && +\frac{P}{2}\sum_m (\cra{2m}\cra{2m}\ana{2m+1}\ana{2m+1} +\text{H.c.})\:,
\end{eqnarray}
where we assume $\hbar=1$, $\cra{m}$ and $\ana{m}$ are creation and annihilation operators for the photon in $m^{\rm{th}}$ cavity, $\hat{n}_m=\cra{m}\,\ana{m}$ is local photon number operator, $\omega_0$ is a cavity eigenfrequency and $J$ is photon tunneling amplitude. Term $\propto U$ is a standard term of Bose-Hubbard Hamiltonian  describing local photon-photon interaction mediated by the nonlinearity of the medium, whereas an extra term $\propto P$ captures direct two-photon hopping. The latter two terms, obviously, do not come into play provided single-particle dynamics is studied, and hence no single-photon topological states are expected.

What is more remarkable, however, is the two-particle sector of this Hamiltonian. As it is straightforward to verify, the Hamiltonian Eq.~\eqref{Hamiltonian} conserves the number of particles and thus the two-photon wave function can be searched in the form
\begin{equation}
 \label{WaveFunc}
 \ket{\psi}=\frac{1}{\sqrt{2}}\,\sum_{m,n} \beta_{mn}\, \cra{m}\,\cra{n}\,\ket{0}
\end{equation}
with the usual normalization $\skvv{\psi}{\psi}=1$ and unknown superposition coefficients $\beta_{mn}$. As a consequence of bosonic symmetry, $\beta_{mn}=\beta_{nm}$ for any indices $m$ and $n$. Inserting Eqs.~\eqref{Hamiltonian}, \eqref{WaveFunc} into the Schr{\"o}dinger equation $\hat{H}\,\ket{\psi}=(\eps+2\,\omega_0)\,\ket{\psi}$ with  $2\,\omega_0$ used as an energy reference, we derive the linear system of equations:
\begin{gather}
  (\eps-2U)\beta_{2m,2m}=-2J \beta_{2m+1,2m}\nonumber\\
    -2J \beta_{2m,2m-1} + P \beta_{2m+1,2m+1}\:,\label{syst1}\\
    (\eps-2U)\beta_{2m+1,2m+1} =-2J\beta_{2m+2,2m+1} \nonumber\\
    -2J\beta_{2m+1,2m}+ P \beta_{2m,2m}\:,\label{syst2}\\
  \eps\beta_{m,n} = - J\beta_{m+1,n} - J \beta_{m-1,n}\notag\\
    - J \beta_{m,n+1} - J \beta_{m,n-1}\mspace{12mu} (m\not=n)\:.\label{syst3}
\end{gather}
In the case of a finite array of length $N$ we additionally impose open boundary conditions $\beta_{00}=\beta_{m0}=0$ and $\beta_{N+1,N+1}=\beta_{m,N+1}=0$ with $m=1,2,\dots N$. As has been pointed out in Refs.~\cite{Longhi,DiLiberto,Gorlach-2017}, these equations can be reinterpreted as an eigenvalue problem for the single particle in a two-dimensional tight-binding lattice. In the latter model, photon-photon interactions $U$ are emulated by the detuning of resonance frequency for the diagonal cavities, whereas the two-photon hopping $P$ is represented as an additional coupling between the diagonal sites.

While the outlined one-dimensional two-particle model can be implemented with optical lattices~\cite{Dutta} or with arrays of transmon qubits~\cite{Roushan:2016NatPhys,Roushan-Science,Ye2019}, the range of parameters attainable in both types of realization is quite limited, the constraints on the magnitude of the direct two-photon hopping being especially strict~\cite{Dutta}. However, in view of the discussed 1D-2D mapping, the same physics can be emulated with two-dimensional classical arrays free of such limitations~\cite{Platero2020}, including, for instance, coupled waveguide lattices~\cite{Mukherjee} or LC circuits~\cite{Olekhno}. For that reason, we examine arbitrary ratios $U/J$ and $P/J$ revealing a full plethora of available effects. The only assumption that is made is the repulsive nature of nonlinearity $U>0$. The spectrum for $U<0$ is immediately recovered by calculating the two-photon states for the system with parameters $-U$ and $-P$ and by inverting the sign of the derived energy. Furthermore, to analyze both possible terminations of the array simultaneously, we focus our attention on the case of odd $N$ when the array starts and terminates with different tunneling links.

To grasp the main features of the proposed system, we start from a simplified model valid in the limit $U\gg J$. In such {\it strong interaction limit} the doublons are tightly bound, i.e. $\beta_{mm}$ coefficients are the dominant ones in the expansion Eq.~\eqref{WaveFunc}. As such, we can rewrite the system Eqs.~\eqref{syst1}-\eqref{syst3} in terms of $\beta_{mm}$ coefficients treating $\beta_{m+1,m}$ as perturbation and fully neglecting the coefficients $\beta_{m+p,m}$ for $p\geq 2$. This approach yields the problem:
\begin{gather}
(\eps-2U-2j)\beta_{2m,2m} =j \beta_{2m-1,2m-1}\nonumber\\
 + (j+P)\beta_{2m+1,2m+1}\:,\label{systeff}\\
(\eps-2U-2j)\beta_{2m+1,2m+1} =(j+P) \beta_{2m,2m} \nonumber\\
 + j \beta_{2m+2,2m+2}\:\label{systeff2}
\end{gather}
with the boundary conditions
\begin{gather}
(\eps-2U-j)\beta_{11} =j \beta_{22}\:,\label{boundeff}\\
(\eps-2U-j)\beta_{N,N} = (j+P)\beta_{N-1,N-1}\:,\label{boundeff2}
\end{gather}
where $j = J^2/U$ is the effective doublon hopping rate associated with two consecutive single-particle tunnelings to the neighboring cavity.

Equations~\eqref{systeff}, \eqref{systeff2} suggest that in the strong interaction limit the dynamics of a doublon is governed by the Su-Schrieffer-Heeger Hamiltonian~\cite{Su} as illustrated by Fig.~\ref{fig:Sketch}(b). This model is known to give rise to the two topologically nontrivial bands with the dispersion
\begin{equation}\label{effsol}
\eps_{\pm}(k) = 2U + 2j \pm \sqrt{j^2 + (j+P)^2 +2j(j+P)\cos{2k}}\:.
\end{equation}
Here Bloch wave number $k$ is defined such that the first Brillouin zone spans the range $[-\pi/2, \pi/2]$. According to Eq.~\eqref{effsol}, the bandgap closing occurs for $|j+P|=|j|$, since this condition renders two tunneling amplitudes equal. Specifically, for $P=0$ and 
\begin{equation}\label{GapClosingSimple}
P=-2J^2/U
\end{equation}
the bandgap closes at $k=\pm\pi/2$ and $k=0$, respectively.

Furthermore, under a suitable parameter choice, doublon bands can be made dispersionless. As a condition for the flat band, we require that $\eps_{+}(0)=\eps_{+}(\pi/2)$, i.e. $|2j+P|=|P|$, which can happen in two situations: (i) trivial case $P\gg j$ or $U\,P\gg J^2$, which implies both strong photon-photon interaction and strong two-photon hopping and (ii) nontrivial case when $j+P=0$ or
\begin{equation}\label{FlatBandSimple}
P=-J^2/U\:.
\end{equation}
In the latter case half of tunneling links in the array vanishes, turning it into the collection of uncoupled dimers.

Besides the intuition about the properties of bulk doublon bands, the developed model also provides some insights into the properties of edge and interface states.  While in the canonical SSH model the edge state arises at the center-of-bandgap frequency being localized at the weak link edge, this case is a bit different because of the interaction-induced detuning of the edge sites by $j$ as suggested by Eqs.~\eqref{boundeff}, \eqref{boundeff2}.

Solving Eq.~\eqref{boundeff2} together with Eqs.~\eqref{systeff}, \eqref{systeff2} for $N\gg 1$, we do not find any localized states near the site $(N,N)$. At the same time, similar analysis for $(1,1)$ site yields two states with degree of localization given by
\begin{equation}\label{LocLeftEdge}
z_{1,2}=e^{2ik}=\frac{j+P}{2\,j^3}\,\left[2j\,P+P^2\pm\sqrt{(2jP+P^2)^2+4\,j^4}\right]\:,
\end{equation}
where localized states correspond to $|z|<1$. The energies of the edge states read:
\begin{equation}\label{EnLeftEdge}
\eps_{1,2}=2U+j-\frac{1}{2j}\,\left[2jP+P^2\pm\sqrt{(2jP+P^2)^2+4j^4}\right]\:.
\end{equation}
Equation~\eqref{LocLeftEdge} shows that higher-energy state $\eps_2$ is localized for any $P\not=0$, while the lower-energy state $\eps_1$ is possible provided
\begin{equation}\label{LocCondition}
-\frac{2\,J^2}{U}<P<0\:.
\end{equation}
Equation~\eqref{LocCondition} is equivalent to the condition $j>|j+P|$, which guarantees that $(1,1)$ site is the strong link edge. Hence, for parameter values given by Eq.~\eqref{LocCondition}, both of the edge states are Tamm-like. In the opposite case $P>0$ or $P<-2J^2/U$ site $(1,1)$ becomes a weak link edge and supports a single state with energy $\eps_2$. Thus, for $P$ outside of the interval Eq.~\eqref{LocCondition} the state $\eps_2$ is a topological one, transforming to the Tamm-like state when the condition Eq.~\eqref{LocCondition} is fulfilled. State $\eps_1$ is a pure Tamm state. Note also that the boundaries of the interval in Eq.~\eqref{LocCondition} coincide with the points of closing and reopening of a bandgap between two doublon bands which illustrates the bulk-boundary correspondence for two-photon topological states in the strong interaction limit.

To further exemplify topological two-photon states, we analyze interface states localized at the boundary of two one-dimensional  arrays with opposite dimerizations. If, for instance, $0^{\rm{th}}$ site is connected with the $1^{\rm{st}}$ and $-1^{\rm{st}}$ sites via the tunneling link $J$, the interface condition in the effective model takes the form:
\begin{equation}\label{IntCond}
\left(\eps-2U-2j\right)\,\beta_{00}=j\,\left(\beta_{11}+\beta_{-1,-1}\right)\:.
\end{equation} 
Hence, the interface site is not detuned with respect to the bulk ones and as a consequence the topological interface state is located exactly in the middle of bandgap $\eps_{\rm{int}}=2U+2j$. If additionally $j>|j+P|$ (short-short defect case), the topological state is also accompanied by two trivial modes lying outside of doublon bandgap~\cite{Blanco-PRL}.

The developed model is only valid in the limit of $U\gg J$. In the next Sec.~\ref{sec:Dispersion} we derive a rigorous solution for the dispersion of bound photon pairs based on Bethe ansatz method and capture a range of intriguing phenomena beyond the canonical SSH model including the interaction of doublon bands with the continuum of scattering states.

\section{Dispersion of bulk doublons}\label{sec:Dispersion}

To solve an infinite set of equations~\eqref{syst1}-\eqref{syst3} and extract the dispersion of photon pairs, one needs some analytic expression for $\beta_{mn}$ coefficients. A powerful approach to this problem is provided by Bethe ansatz technique~\cite{Essler,Karbach}. The standard Bethe ansatz has the form:
\begin{equation}\label{eq:Bethe0}
\beta_{mn}=C\,\exp\left[i\frac{k}{2}\,(m+n)+i\frac{\kap}{2}(m-n)\right]
\end{equation}
for $m\geq n$. In this expression, $k$ is Bloch wave number describing the motion of photon pair as a whole, whereas $\kap$ captures the relative motion of particles. Bound photon pairs are characterized by complex $\kap$, in which case the wave function decays with the increase of separation $(m-n)$ between the photons.

While such simple ansatz captures the properties of bound pairs in the limiting case $P=0$, it appears to be inconsistent with Eqs.~\eqref{syst1}-\eqref{syst3} in the general case of $P\not=0$ and arbitrary $k$. To proceed with the analytical solution, we need to incorporate into the ansatz the presence of {\it two} sites in the unit cell. This extended unit cell shrinks the first Brillouin zone for doublons from $[-\pi, \pi]$ (as is the case for $P=0$) to $[-\pi/2,\pi/2]$ mixing the states with wave numbers $k$ and $k+\pi$. Therefore, we introduce the following modification of Bethe ansatz:
\begin{gather}
  \beta_{mn}=C_{1}\,e^{ik(m+n)/2}\,e^{i\kap_1(m-n)/2}\nonumber\\
    +C_{2}\,\,e^{i(k+\pi)(m+n)/2}\,e^{i\kap_2(m-n)/2}\label{eq:ansatz}
\end{gather}
with $m\geq n$ and $\text{Im}\,\kap_{1,2}>0$. The modified ansatz Eq.~\eqref{eq:ansatz} appears to be consistent with full system of equations Eqs.~\eqref{syst1}-\eqref{syst3} and determines doublon dispersion as further detailed in Appendix A.

Omitting the details of the derivation, we would like to stress here several simple but illuminating results. The energies of doublon bands in the limiting case $k=\pm\pi/2$ can be found analytically:
\begin{equation}\label{EPi2}
\eps_{\pm}=\sgn{[2U\pm P]}\sqrt{(2U\pm P)^2+8J^2}\:.
\end{equation}
Thus, bound photon pairs are always stable for wave numbers near the boundaries of the first Brillouin zone. In the strong interaction limit, the energies of the two bands scale as $(2U+P)$ and $(2U-P)$, which means that the effective photon-photon interaction $U$ defines the average energy of bound pair, whereas the two-photon hopping $P$ controls energy splitting between the two bands.

For $k=0$, energies of the doublon states read
\begin{gather}
\eps'_+=\sgn{[2U+P]}\sqrt{(2 U+P)^2+16J^2}\:,\\
\eps'_{-}=2U-P\:,
\end{gather}
where $\eps_-$ and $\eps'_-$ ($\eps_+$ and $\eps'_+$) can correspond to the same or to the different doublon bands. Note that the doublon band associated with $\eps'_{-}$ can collapse intersecting with the continuum of two-photon scattering states for nonzero $k$ sufficiently far from the Brillouin zone boundaries. As we show in Appendix A, collapse of the doublon band occurs in the range of parameters
\begin{equation}
-4J<2U-P<4J\:.
\end{equation}

    \begin{center}
    \begin{figure}[b!]
     \includegraphics[width=\linewidth]{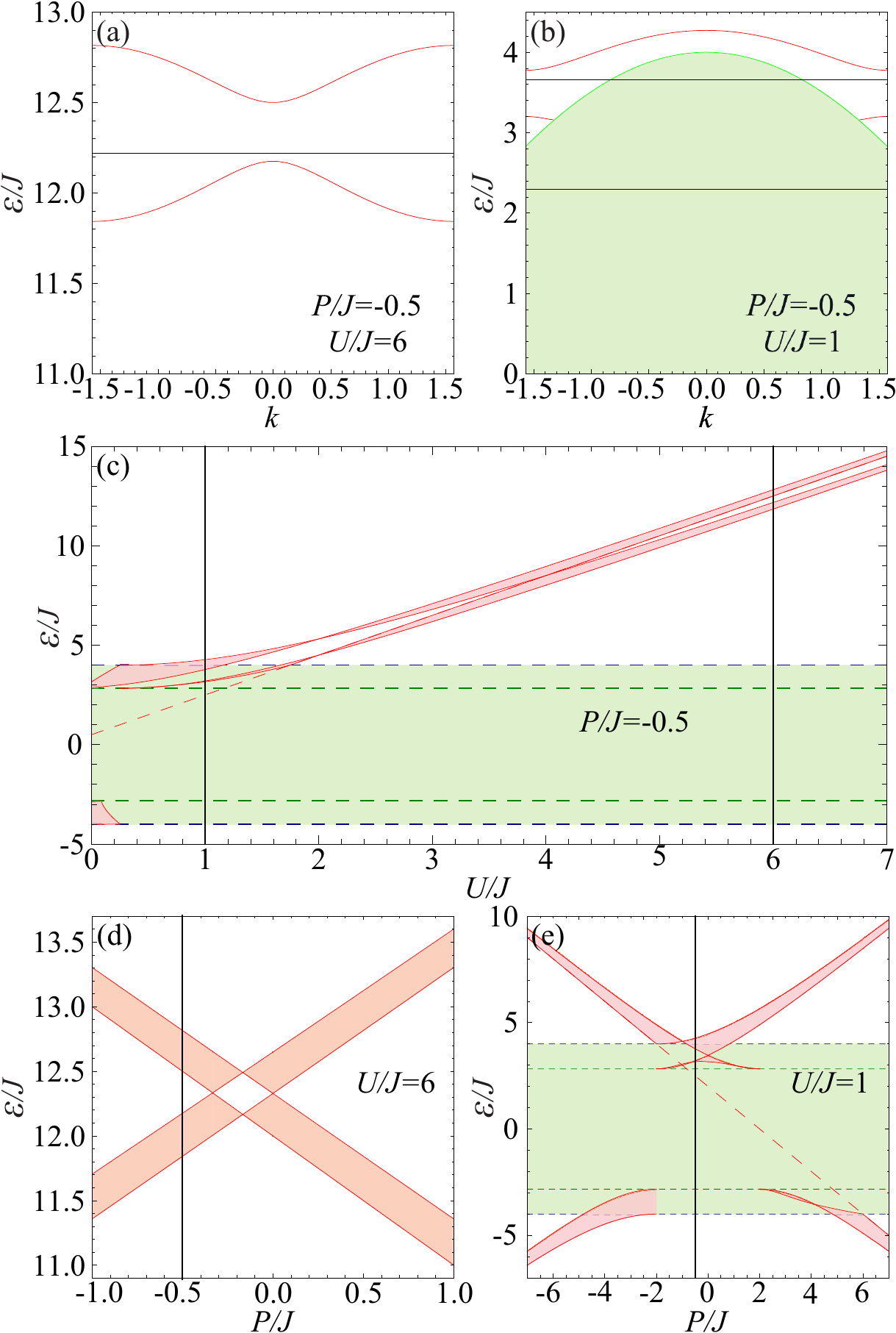}
     \caption{Dispersion of two-photon excitations in the extended Bose-Hubbard model Eq.~\eqref{Hamiltonian}. (a,b) Doublon dispersion for the two representative cases: (a) strong interaction limit $U/J = 6$, $P/J = -0.5$; (b) moderate interactions $U/J = 1$, $P/J = -0.5$, when  lower doublon band intersects with the continuum of scattering states and collapses. Two Tamm-like doublon edge states exist in the scattering continuum.  In both cases, red solid lines correspond to bulk doublons, light green continuum shows energies of two-photon scattering states, horizontal black lines indicate energies of doublon edge states.  (c) Evolution of doublon bands shown by red when two-photon hopping $P/J=-0.5$ is fixed and the interaction strength $U$ is varied. Black vertical lines indicate the parameter values used in panels (a) and (b). The green band shows the range of energies for the two-photon scattering states. (d,e) Evolution of doublon bands when photon-photon interaction $U/J$ is fixed and the two-photon hopping $P/J$ is varied. Vertical lines indicate the magnitude of $P$ corresponding to panels (a,b). (d) Strong interactions  $U/J=6$. (e) Moderate interactions $U/J=1$.
        }
    \label{fig:DoublonZones}
    \end{figure}
    \end{center}

To illustrate the obtained solution further, we explore the dispersion of doublons in two characteristic situations with the same two-photon hopping $P/J=-0.5$ and different magnitude of the effective photon-photon interaction: sufficiently strong interactions $U/J=6$, Fig.~\ref{fig:DoublonZones}(a) and moderate interactions $U/J=1$, Fig.~\ref{fig:DoublonZones}(b). The former case, shown in Fig.~\ref{fig:DoublonZones}(a), exhibits a remarkable agreement with the effective SSH model, both in terms of the boundaries of the bulk bands given by Eq.~\eqref{effsol} and spectral position of topological edge state. However, when the strength of interaction $U$ is decreased, one of the  doublon bands intersects with the continuum of scattering states and collapses [Fig.~\ref{fig:DoublonZones}(b)]. In agreement with our previous analysis, both doublon bands are stable near the edges of the Brillouin zone, while one of the bands becomes unstable in the vicinity of $k=0$. At the same time, the edge state near the first site persists coexisting with the continuum of scattering states.

It is also instructive to trace the evolution of doublon bands when the parameters of the model $U$ and $P$ are varied. Specifically, Fig.~\ref{fig:DoublonZones}(c) illustrates the evolution of doublon bands with the interaction strength $U$ for the fixed value of two-photon tunneling $P$. In accordance with Eq.~\eqref{EPi2}, we observe that the continuum of scattering states can be located right between two doublon bands provided $(2U+P)$ and $(2U-P)$ have different signs. The decreased spectral width of the lower band inside the continuum of scattering states for $0.3<U/J<1.75$ serves as an evidence of doublon collapse. In accordance with the simplified model developed in Sec.~\ref{sec:Model}, doublon bands become dispersionless for $U/J=2$ [cf. Eq.~\eqref{FlatBandSimple}] and the gap between them closes at $U/J=4$ [cf. Eq.~\eqref{GapClosingSimple}]. In fact, such close agreement is not occasional, since the conditions for the flat band and for gap closing predicted by the simplified model coincide with those obtained from the rigorous solution as discussed in Appendix A.

In the strong interaction limit, two-photon hopping $P$ is the only parameter controlling the separation of two doublon bands. Figure~\ref{fig:DoublonZones}(d) shows almost linear dependence of doublon energies on the magnitude of $P$, illustrating topological transitions due to closing and reopening of bandgap and the emergence of flat bands. The situation appears to be more complicated for moderate interactions $U/J=1$, when doublon bands interact with the scattering continuum collapsing and reviving [Fig.~\ref{fig:DoublonZones}(e)]. The doublon state with energy $(2U-P)$ shown by the red dashed line in Fig.~\ref{fig:DoublonZones}(e) appears to be especially robust crossing the entire scattering continuum.

\section{Edge and interface topological doublon states}\label{sec:Edge}

Closing and reopening of a bandgap between two doublon bands demonstrated in Sec.~\ref{sec:Dispersion} hints towards topological transitions happening in the system. While strong interaction limit $U\gg J$ is well-understandable in terms of the effective SSH-type model, the case of moderate interaction appears to be less intuitive. A characteristic example is presented in Fig.~\ref{fig:DoublonZones}(b) when one of the doublon bands partially collapses and the edge state appears in the continuum of scattering states. These observations demonstrate two important features of our system.


    \begin{center}
    \begin{figure}[b!]
     \includegraphics[width=\linewidth]{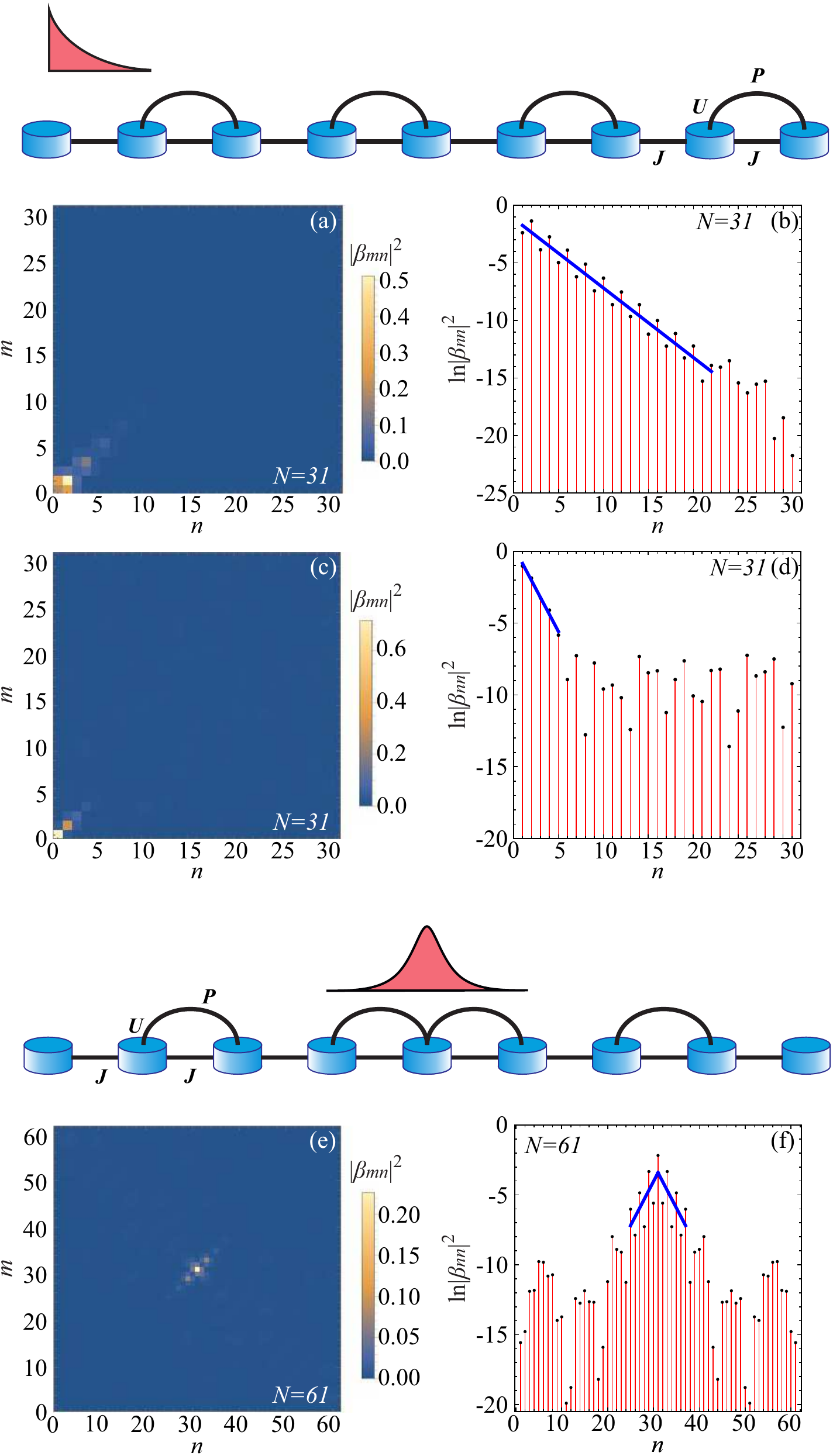}
     \caption{Realization of doublon edge and interface states in the continuum for two geometries illustrated by insets. (a,c) Probability distributions $|\beta_{mn}|^2$ for Tamm-like doublon edge states localized near the first site of the array of $N=31$ cavity with energies $\eps^{(\rm{edge})}/J = 3.66$ and  $\eps^{(\rm{edge})}/J = 2.29$,  respectively. (b,d) Probability distributions $|\beta_{nn}|^2$ versus $n$ in logarithmic scale for the same states as in panels (a,c), respectively. Both edge states feature non-exponential localization. (e) Probability distribution for topological interface state of a doublon with energy  $\eps^{(\rm{int})}/J = 3.53$ localized at the boundary between the two arrays with different dimerizations and overall length of $N = 61$ cavity. (f) Probability distribution $|\beta_{nn}|^2$ versus $n$ in logarithmic scale for the interface state in panel (e). The decay of the interface state with distance is not captured by simple exponential formula. The calculations are performed for $U/J = 1$, $P/J = -0.5$ as in Fig.~\ref{fig:DoublonZones}(b).
       }
    \label{fig:EdgeState}
    \end{figure}
    \end{center}

First, the problem of bulk-boundary correspondence in two-particle topological models becomes more involved, since the corresponding doublon band can collapse leaving Berry connection undefined, whereas the topological state  persists.

Second, the two-particle bound edge state can coexist with the continuum of scattering states as has been previously pointed out for a different two-particle model~\cite{Zhang2012,Zhang2013} thus providing a realization of the two-particle bound state in the continuum (BIC)~\cite{Soljacic-bic}.

To get further intuition about doublon BIC arising in this system, we analyze two geometries: (i) semi-infinite array with the edge state localized near the first site; (ii) domain wall between the two arrays with different dimerizations hosting the interface state. In the first scenario illustrated in Fig.~\ref{fig:EdgeState}(a-d) we observe two edge states of bound photon pairs. The state with higher energy, shown in panels (a,b), still retains quite good localization close to exponential. At the same time, lower-energy state [Fig.~\ref{fig:EdgeState}(c,d)] which lies deeper in the scattering continuum clearly exhibits non-exponential localization caused by the stronger interaction with the continuum. In turn, the interface state Fig.~\ref{fig:EdgeState}(e) exhibits a symmetric profile with respect to the domain wall thus resembling the interface state at a long-long defect in the canonical SSH model~\cite{Blanco-PRL}. However, in contrast to the SSH case, the localization of the interface state is non-exponential [Fig.~\ref{fig:EdgeState}(f)] due to its hybridization with the two-photon scattering states.

\section{Discussion and conclusions}\label{sec:Concl}

To summarize, we have investigated a system with interaction-induced topological order. Even though the single-particle model is topologically trivial, the two-particle bands feature a topological bandgap with doublon edge and interface states inside it. Quite interestingly, the observed two-particle topological states remain stable under the collapse of the bulk doublon band and, moreover, they can coexist with the scattering continuum thus providing a realization of the two-photon interaction-induced BIC.

\begin{center}
    \begin{figure}[h!]
     \includegraphics[width=0.65\linewidth]{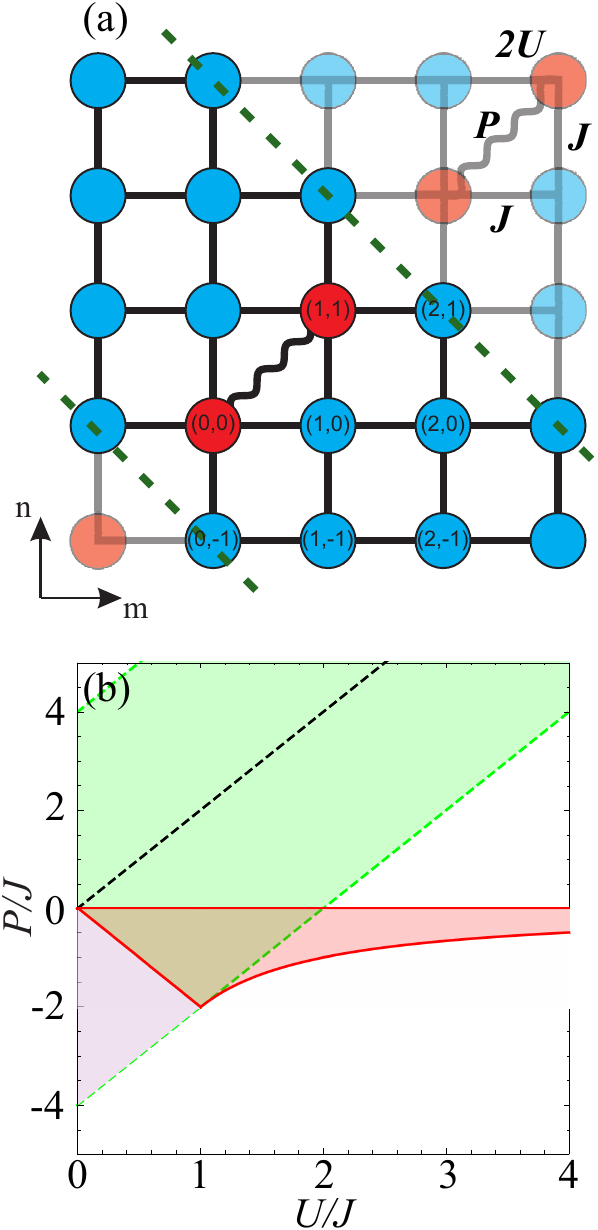}
     \caption{(a) Illustration of various two-photon Fock states $\cra{m}\,\cra{n}\,\ket{0}$ comprising the two-photon wave function on a two-dimensional map. Boundaries of the inversion-symmetric unit cell used for the Zak phase calculation for doublon bands are shown by green dashed lines. (b) Phase diagram showing the magnitude of the Zak phase as a function of model parameters $U/J$ and $P/J$. Zak phase is equal to $\pi$ in the red shaded area of the diagram and equal to $0$ elsewhere. Areas shaded by light green and light purple indicate the parameter ranges corresponding to the collapse of lower and upper doublon bands, respectively. The black dashed line corresponds to the condition $P=2U$.
}
    \label{fig:PhaseDiagram}
   \end{figure}
\end{center}

To prove the topological origin of the observed doublon edge states, we perform the calculation of the Zak phase for bulk doublon bands. Our results shown in Fig.~\ref{fig:PhaseDiagram} and summarized in Appendix B suggest extremely rich physics associated with topological transitions, collapse and revival of doublon bands and edge states taking place as the parameters of the system are varied.

We believe that the physical realization of the proposed model can be based on cold atomic gases in optical lattices, arrays of coupled transmon qubits or other systems featuring a significant anharmonicity of on-site potential. At the same time, mapping of interacting one-dimensional two-body problem onto the two-dimensional classical setup possible for the class of Bose-Hubbard models opens another experimentally feasible route to emulate topological states in interacting systems.

\section*{Acknowledgments}

We acknowledge valuable discussions with Alexander Poddubny, Nikita Olekhno and Marco Di Liberto. This work was supported by the Russian Science Foundation (grant No.~16-19-10538). A.A.S. acknowledges partial support by Quantum Technologies Center, Faculty of Physics, Lomonosov Moscow State University.  M.A.G. acknowledges partial support by the Foundation for the Advancement of Theoretical Physics and Mathematics ``Basis".

\section*{Appendix A. Calculation of bulk doublon dispersion}

In this Appendix, we outline the rigorous solution for the dispersion of bulk doublons based on modified Bethe ansatz, Eq.~\eqref{eq:ansatz}. We seek the solution of equations
\begin{gather}
  \eps\beta_{m,n} = - J\beta_{m+1,n} - J \beta_{m-1,n}\nonumber\\
    - J \beta_{m,n+1} - J \beta_{m,n-1}\mspace{12mu} (m\not=n),\label{sys1}\\
  (\eps-2U)\beta_{2m,2m}=-2J \beta_{2m+1,2m}\nonumber\\
    -2J \beta_{2m,2m-1} + P \beta_{2m+1,2m+1}\:,\label{sys2}\\
    (\eps-2U)\beta_{2m+1,2m+1} =-2J\beta_{2m+2,2m+1} \nonumber\\
    -2J\beta_{2m+1,2m}+ P \beta_{2m,2m}\label{sys3}
\end{gather}
assuming that
\begin{equation}\label{Bethe}
\begin{split}
  \beta_{mn}=C_{1}\,e^{ik(m+n)/2}\,e^{i\kap_1(m-n)/2}\\
    +C_{2}\,\,e^{i(k+\pi)(m+n)/2}\,e^{i\kap_2(m-n)/2}
\end{split}
\end{equation}
for any $m\geq n$ and $\Im\kap_{1,2}>0$. Inserting Eq.~\eqref{Bethe} into the system Eqs.~\eqref{sys1}-\eqref{sys3}, we get:
\begin{gather}
\eps=-4J\,\cos\frac{k}{2}\,\cos\frac{\kap_1}{2}\:,\label{res1}\\
\eps=4J\,\sin\frac{k}{2}\,\cos\frac{\kap_2}{2}\:,\label{res2}\\
C_1\,\left[\eps-2U+4J\,\cos\frac{k}{2}\,e^{i\kap_1/2}-P\,e^{ik}\right]\notag\\
+C_2\,\left[\eps-2U-4J\,\sin\frac{k}{2}\,e^{i\kap_2/2}+P\,e^{ik}\right]=0\:,\label{res3}\\
C_1\,\left[\eps-2U+4J\,\cos\frac{k}{2}\,e^{i\kap_1/2}-P\,e^{-ik}\right]\notag\\
+C_2\,\left[\eps-2U-4J\,\sin\frac{k}{2}\,e^{i\kap_2/2}+P\,e^{-ik}\right]=0\:.\label{res4}
\end{gather}
Here, Eqs.~\eqref{res1}-\eqref{res2} are obtained from the single Eq.~\eqref{sys1} due to the linear independence of $e^{i\kap_1\,(m-n)/2}$ and $e^{i\kap_2\,(m-n)/2}$ for $m\not=n$. The system of four equations \eqref{res1}-\eqref{res4} defines four unknowns: $\eps$, $\kap_1$, $\kap_2$ and the ratio $C_1/C_2$, while the absolute values of $C_1$ and $C_2$ are determined from the normalization of the two-photon wave function.

Excluding doublon energy from Eqs.~\eqref{res3}-\eqref{res4} with Eqs.~\eqref{res1}-\eqref{res2} and further rearranging them, we arrive to the linear system with respect to $C_1$ and $C_2$:
\begin{align}
& C_1\,\left[4iJ\,\cos\frac{k}{2}\,\sin\frac{\kap_1}{2}-2U-P\,\cos k\right]\nonumber\\
& +iC_2\,P\,\sin k=0\:,\label{lin1}\\
&-iC_1\,P\,\sin k\nonumber\\
&+C_2\,\left[-4iJ\,\sin\frac{k}{2}\,\sin\frac{\kap_2}{2}-2U+P\,\cos k\right]=0\:.\label{lin2}
\end{align} 
Setting the determinant of this system to zero, we recover that
\begin{gather}
8J^2\,\sin k\,\sin\frac{\kap_1}{2}\,\sin\frac{\kap_2}{2}-4iJ\,\cos\frac{k}{2}\,\sin\frac{\kap_1}{2}\,(2U-P\,\cos k)\nonumber\\
+4iJ\,\sin\frac{k}{2}\,\sin\frac{\kap_2}{2}\,\left(2U+P\,\cos k\right)+4\,U^2-P^2=0\:.\label{LocCond}
\end{gather}
To provide an efficient numerical algorithm to calculate the dispersion of doublons, we introduce two auxiliary dimensionless variables:
\begin{align}
x\equiv -i\cos\frac{k}{2}\,\sin\frac{\kap_1}{2}\:,\label{xdef}\\
y\equiv -i\sin\frac{k}{2}\,\sin\frac{\kap_2}{2}\:.\label{ydef}
\end{align}
Making use of Eqs.~\eqref{res1}, \eqref{res2}, \eqref{LocCond}, we get the following closed-form system of equations:
\begin{align}
& f(x,y,k)\equiv x^2-y^2+\cos k=0\:,\label{num1}\\
& g(x,y,k)\equiv -16J^2\,x\,y+4Jx\,(2U-P\,\cos k)\nonumber\\
&-4Jy\,(2U+P\,\cos k)+4U^2-P^2=0\:.\label{num2}
\end{align}
Separating real and imaginary parts for $\kap_1$ and $\kap_2$ and taking into account that $\Im\kap_{1,2}>0$, we can show that the real parts of $x$, $y$ and $\eps$ satisfy the following condition:
\begin{equation}\label{SignCond}
\sgn\eps=-\sgn\,x=\sgn\,y\:.
\end{equation}
Hence, to find doublon dispersion one has to solve the system of algebraic equations \eqref{num1}-\eqref{num2} numerically, keeping only those pairs of $x$ and $y$ which satisfy an additional constraint Eq.~\eqref{SignCond}. Doublon energy is then given by
\begin{equation}
\eps=-4J\,\sgn x\,\sqrt{\cos^2\frac{k}{2}+x^2}\:.
\end{equation}

Note that bound pairs coexist with the continuum of two-photon scattering states characterized by real $k$ and $\kap$. Bulk dispersion of such states is captured by the simple formula~\eqref{res1}, where both parameters are real.

In some limiting cases, the expression for the doublon energy simplifies considerably. For instance, if $k=\pm\pi/2$, $x=-(2U\pm P)/(4J)$, whereas doublon energy
\begin{equation}\label{RootPi}
\eps_{\pm}=\sgn\left(2U\pm P\right)\,\sqrt{(2U\pm P)^2+8\,J^2}\:.
\end{equation}
Hence, doublons with $k$ at the edge of Brillouin zone are always stable.

Another limiting case is $k=0$ when one of the doublon bands has $C_2=0$ and Bethe ansatz in the standard form Eq.~\eqref{eq:Bethe0} can be applied. In such a case, the energy of this particular doublon band is given by
\begin{equation}\label{Root01}
\eps'_{+}=\sgn(2U+P)\,\sqrt{(2U+P)^2+16J^2}\:,
\end{equation}
whereas the second doublon solution is localized purely on the diagonal and has the energy
\begin{equation}\label{Root02}
\eps'_{-}=2U-P\:.
\end{equation}
Note, however, that the latter doublon band is unstable for nonzero $k$ far enough from the Brillouin zone boundaries.

With the obtained rigorous solution, we can also find the condition for closing of the gap between two doublon bands. To this end, two roots for $\eps(k=0)$ given by Eqs.~\eqref{Root01} and \eqref{Root02} should coincide, i.e. 
\begin{equation}\label{GapClosing}
UP=-2J^2\:.
\end{equation}

In a similar manner we examine the condition for flat band. Inspecting Eqs.~\eqref{RootPi}, \eqref{Root01}, \eqref{Root02}, we find out that $\eps_{+}=\eps'_{-}$ and $\eps_{-}=\eps'_{+}$ provided
\begin{equation}\label{FlatBand}
UP=-J^2\:.
\end{equation}
Note that $\eps_{+}$ and $\eps'_{+}$ can correspond to the same or to the different doublon bands depending on parameter choice, and the subscript $\pm$ is used here just to label the solutions.

Quite interestingly, both of these results, Eqs.~\eqref{GapClosing} and \eqref{FlatBand} coincide with those obtained from the simplified SSH-type effective model, Eqs.~\eqref{GapClosingSimple}, \eqref{FlatBandSimple}. 
Additionally, we can find the conditions for collapse of the doublon state. Collapsing band is characterized by the  decreased co-localization of photons, i.e. $x\rightarrow 0$. Solving Eqs.~\eqref{num1}, \eqref{num2} for $k=0$ with $x=0$, we find the following condition:
\begin{equation}\label{CollapseCond}
2U-P=\pm 4J\:.
\end{equation}
In a similar manner one can also examine the collapse of doublon with arbitrary wave number $k$.

\section*{Appendix B. Calculation of the Zak phase for bulk doublon bands}

An obtained analytic solution for the dispersion of doublons allows us to evaluate the Zak phase for bulk doublon bands. To this end, we take the periodic part $\ket{u_k}$ of the full doublon wave function in the following form: 

\begin{equation}
\begin{split}
  \label{WaveFuncCell}
  \ket{u_k} =\beta_{0,0}\ket{2_0}+\beta_{1,1}\ket{2_1}+\sqrt{2}\sum_{n=1}^{\infty} \beta_{n,-n} \ket{1_{n}1_{-n}}\\+\sqrt{2}\sum_{n=2}^{\infty} \beta_{n,2-n} \ket{1_{n}1_{2-n}}
+\dfrac{1}{\sqrt{2}}\sum_{n=1}^{\infty} \beta_{n-1,-n} \ket{1_{n-1}1_{-n}}\\+\sqrt{2}\sum_{n=1}^{\infty} \beta_{n,1-n} \ket{1_{n}1_{1-n}}+\dfrac{1}{\sqrt{2}}\sum_{n=1}^{\infty} \beta_{n+1,2-n} \ket{1_{n+1}1_{2-n}}\:,
\end{split}
\end{equation}
where $\ket{1_m 1_n}\equiv \cra{m}\,\cra{n}\,\ket{0}$ for $m\not=n$ and $\ket{2_m}\equiv 2^{-1/2}\,\cra{m}\,\ket{0}$. The corresponding choice of the unit cell is illustrated in Fig.~\ref{fig:PhaseDiagram}(a). Importantly, such unit cell is inversion-symmetric which ensures quantization of the Zak phase in units of $\pi$~\cite{Zak}.

In turn, the coefficients $\beta_{mn}$ are defined from Eq.~\eqref{Bethe} and hence
\begin{eqnarray}
\ket{u_k}&=&C_1\ket{v_1} +C_2\ket{v_2}\:,\\
\ket{v_1}&=& \ket{2_0} +e^{ik}\ket{2_1}+\sqrt{2}\sum_{n=1}^{\infty} e^{i\kap_1 n} \ket{1_{n}1_{-n}}\nonumber\\&&
+\sqrt{2}\sum_{n=2}^{\infty} e^{i\kap_1(n-1)} e^{i k} \ket{1_{n}1_{2-n}}\nonumber\\
&&+\dfrac{1}{\sqrt{2}}\sum_{n=1}^{\infty}e^{i\kap_1(n-\frac{1}{2})} \bigg{[} e^{-\frac{ik}{2}}\ket{1_{n-1}1_{-n}}\nonumber\\&&+2e^{\frac{ik}{2}} \ket{1_{n}1_{1-n}}+e^{\frac{3ik}{2}} \ket{1_{n+1}1_{2-n}}\bigg{]}\:,\label{eq:v1}
\end{eqnarray}
\begin{eqnarray}
\ket{v_2}&=& \ket{2_0} - e^{ik}\ket{2_1}+\sqrt{2}\sum_{n=1}^{\infty} e^{i\kap_2 n} \ket{1_{n}1_{-n}}\nonumber\\&&
- \sqrt{2}\sum_{n=2}^{\infty} e^{i\kap_2(n-1)}e^{ik} \ket{1_{n}1_{2-n}}\nonumber\\
&&+\dfrac{1}{\sqrt{2}}\sum_{n=1}^{\infty}e^{i\kap_2(n-\frac{1}{2})} \bigg{[} -i e^{-\frac{ik}{2}}\ket{1_{n-1}1_{-n}}\nonumber\\&&+2 i e^{\frac{ik}{2}} \ket{1_{n}1_{1-n}} - i e^{\frac{3ik}{2}} \ket{1_{n+1}1_{2-n}}\bigg{]}\:.\label{eq:v2}
\end{eqnarray}

The Zak phase is defined in terms of Berry connection as
\begin{equation}
\gamma=\int\limits_{-\pi/2}^{\pi/2}\,A(k)\,dk\:,
\end{equation}
where Berry connection $A(k)$
\begin{eqnarray}\label{eq:BerryConnection}
A(k)&=& i \bigg{\langle} u_k\ket{\frac{\partial u_k}{\partial k}} 
=i C_1^*\frac{\partial C_1}{\partial k}\langle v_1 \ket{v_1} 
+i C_1^*\frac{\partial C_2}{\partial k}\langle v_1 \ket{v_2}\nonumber\\&& 
+i C_1^*C_1 \bigg{\langle}  v_1 \ket{\frac{\partial v_1}{\partial k}}
+i C_1^*C_2 \bigg{\langle}  v_1 \ket{\frac{\partial v_2}{\partial k}} \nonumber\\
&&+i C_2^*\frac{\partial C_1}{\partial k}\langle v_2 \ket{v_1} 
+i C_2^*\frac{\partial C_2}{\partial k}\langle v_2 \ket{v_2}\nonumber\\&&
+i C_2^*C_1 \bigg{\langle}  v_2 \ket{\frac{\partial v_1}{\partial k}}
+i C_2^*C_2 \bigg{\langle}  v_2 \ket{\frac{\partial v_2}{\partial k}} \:.
\end{eqnarray}
Using Eqs.~\eqref{eq:v1}-\eqref{eq:v2}, we calculate the scalar products:
\begin{eqnarray}
\langle v_1\ket{v_1}&=& \frac{2(1+x_{11})+3\sqrt{x_{11}}}{1-x_{11}}\:, \\
\langle v_1\ket{v_2}&=&\langle v_2\ket{v_1}^*= i\frac{\sqrt{x_{12}}}{1-x_{12}}\:,\\
\langle v_2\ket{v_2}&=& \frac{2(1+x_{22})+3\sqrt{x_{22}}}{1-x_{22}}\:,
\end{eqnarray}

\begin{eqnarray}
\bigg{\langle}  v_1\ket{\frac{\partial v_1}{\partial k}}&=&i\kap_1'\frac{8x_{11}+3\sqrt{x_{11}}(1+x_{11})}{2(1-x_{11})^2}\nonumber\\&&
+ i\frac{2(1+x_{11})+3\sqrt{x_{11}}}{2(1-x_{11})}\:, \\
\bigg{\langle}  v_1\ket{\frac{\partial v_2}{\partial k}}&=&-\frac{2i(1+x_{12})
+\sqrt{x_{12}}}{2(1-x_{12})}\nonumber\\&&
-\kap_2'\frac{\sqrt{x_{12}}(1+x_{12})}{2(1-x_{12})^2}\:,\\
\bigg{\langle}  v_2\ket{\frac{\partial v_1}{\partial k}}&=&-\frac{2i(1+x_{21}) -\sqrt{x_{12}}}{2(1-x_{21})}\nonumber\\&&
+\kap_1'\frac{\sqrt{x_{21}}(1+x_{21})}{2(1-x_{21})^2} \:,\\
\bigg{\langle}  v_2\ket{\frac{\partial v_2}{\partial k}}&=&i\kap_2'\frac{8x_{22}+3\sqrt{x_{22}}(1+x_{22})}{2(1-x_{22})^2}\nonumber\\&&
+i\frac{2(1+x_{22})+3\sqrt{x_{22}}}{2(1-x_{22})}\:,
\end{eqnarray}
where $x_{\alpha\beta} = (e^{i\kap_\alpha})^*e^{i\kap_\beta}$.

The scalar products $\bra{\dfrac{\partial v_i}{\partial k}}v_j\bigg{\rangle} $ are obtained by complex conjugation of $\bigg{\langle}  v_j\ket{\dfrac{\partial v_i}{\partial k}}$. Calculating Berry connection, we need to determine five quantities: \Bigg{(}$\dfrac{\partial C_1}{\partial k}$, $\dfrac{\partial C_2}{\partial k}$, $\dfrac{\partial \kap_1}{\partial k}$, $\dfrac{\partial \kap_2}{\partial k}$, $\dfrac{\partial \eps}{\partial k}$\Bigg{)}. Differentiation of the identities Eqs. \eqref{res1}-\eqref{res4} with respect to $k$ yields only four equations. One more equation can be obtained differentiating the identity 
\begin{equation}\label{NormCond}
\langle u_k\ket{u_k} =1\:,
\end{equation}
which is the normalization condition for the periodic part of the wave function. In these calculations, the gauge of the wave function should be fixed. To ensure smooth behavior of $C_1$ and $C_2$ coefficients with $k$, we choose their phases such that  $(C_1+C_2) e^{i\phi}$ is real, where:
\begin{eqnarray}
e^{i\phi} &=& \sqrt{\dfrac{-P e^{-ik}-2U+4iJ\sin{\kap_1/2}\cos{k/2}}{P e^{ik}+2U-4iJ\sin{\kap_1/2}\cos{k/2}}}\:.
\end{eqnarray}
Such choice ensures, in particular, that the Berry connection is a smooth function of wave number $k$. The Zak phase calculation is then accomplished in several steps:

1) Complex coefficients $C_{1,2}$ are calculated using normalization condition Eq.~\eqref{NormCond} and Eqs.~\eqref{lin1}, \eqref{lin2}.

2) The derivatives $\dfrac{\partial C_1}{\partial k}$,$\dfrac{\partial C_2}{\partial k}$, $\dfrac{\partial \kap_1}{\partial k}$, $\dfrac{\partial \kap_2}{\partial k}$, $\dfrac{\partial \eps}{\partial k}$ are evaluated from the differentiated identities Eqs.~\eqref{res1}-\eqref{res4} and \eqref{NormCond}.

3) The obtained quantities are inserted into Eq.~\eqref{eq:BerryConnection} and Berry connection is evaluated numerically in the entire Brillouin zone. The Zak phase is recovered by numerical integration.

Performing the calculation of the Zak phase for different values of model parameters, $U$ and $P$, we plot the phase diagram shown in Fig.~\ref{fig:PhaseDiagram}(b) and analyze topological transitions happening in the system. For unit cell choice shown in Fig.~\ref{fig:PhaseDiagram}(a), the link with the direct two-photon hopping is located inside the unit cell. For $U>1$, $\gamma=\pi$ is achieved exactly in the same range of parameters when $|j+P|<j$, i.e. the weak tunneling link appears {\it inside} the unit cell. This is consistent with the result expected from the effective SSH model with coupling constants equal to $j=J^2/U$ and $j+P$.

However, besides the analogy with the effective SSH model, our system also exhibits some distinctive properties. First, even if one of the doublon bands collapses, the Zak phase can still be defined using the wave functions for the remaining band. Furthermore, despite the collapse of the bulk doublon band, the doublon edge state can persist, which provides an interesting feature of two-photon topological states.

Even though this model is just a particular example, we believe that the present analysis provides valuable insights into the topological properties and bulk-boundary correspondence in nonlinear topological models.

\bibliography{TopologicalLib}

\end{document}